# Novel Time-Independent Model for Determining the Critical Mass of a Fissile Nuclide

B. Cameron Reed

Department of Physics

Alma College

Alma, MI 48801

Ph: (989) 463-7266

Fax: (989) 463-7076

e-mail: reed@alma.edu

**Abstract**

A straightforward, time-independent model for determining the critical mass of a spherical sample of a fissile nuclide is developed from basic neutron collision theory and elementary probability arguments. This model is suitable for presentation to even undergraduate students and gives results in excellent accord with published values.

**1.     Introduction**

A central idea involved with nuclear weapons is that there exists a necessary "critical mass" of fissile material. However, formal derivations of expressions for determining critical masses involve time-dependent diffusion theory [1-3], material which may be unfamiliar to readers desiring a straightforward analysis of the situation. To this end, development of a simple but physically realistic time-independent model for estimating critical masses would appear to be worthwhile. This paper presents such a model, which, despite its limitations, gives results in excellent agreement with published values.

Qualitatively, the concept of a critical mass can be easily expressed. Some species of nuclei fission upon being struck by a bombarding neutron, consequently releasing *secondary* neutrons. The average number of secondary neutrons emitted per fission is designated ν. These secondary neutrons can potentially go on to induce other fissions, resulting in a chain reaction. However, a well-known law from reactor theory demands that a certain number of neutrons must reach the surface of the sample and escape. The number that do so depends upon the size of the sample and factors such as its density and the propensity for the nuclei to fission upon being struck. If the sample is physically small most of the neutrons will escape. *When the sample is just large enough so that those neutrons that do not escape number just enough to induce fission in all nuclei, one has a critical mass.*

Let the total number of nuclei in the sample be N. If all are to fission, then νN secondary neutrons will ultimately be created. (It is customary to imagine having available a few "initiator" neutrons to start the reaction; all but a handful of nuclei are fissioned by the action of secondary neutrons.) All νN secondary neutrons will thus either be consumed in fissions or escape: $N_{created}$

= $N_{escape} + N_{fission}$. Since the condition to have a critical mass is that each nucleus does fission, then N neutrons must be consumed in fissions, that is, $\nu N = N_{escape} + N$. Hence, we can *permit* (ν-1)N neutrons to escape. However, the law alluded to above dictates how many *must* escape. The condition for a critical mass is then to have things so that the number that must escape is less than or equal to the number that can be permitted to escape:

$$N_{escape} \leq (\nu-1)N. \qquad (1)$$

The essence of determining critical mass is to develop an expression for the left side of Eq. (1) via the escape law alluded to above. This is done in Section 2 below for a spherical sample of fissile material; the result proves to involve a (yet unknown) critical radius R in such a way that an analytic solution is impossible. Numerical solution techniques are discussed in Section 3. Results for a number of nuclides are presented in Section 4, along with a brief discussion of how one can in practice lower the critical mass from what is calculated. A brief summary is given in Section 5.

**2.    Cross-sections and escape probabilities**

The probability of a given reaction occurring when an incident particle strikes a target nucleus is quantified by way of a *cross-section* σ for that reaction. This can be thought of as giving an effective area that a nucleus presents to a bombarding particle for the reaction of interest. The standard unit for cross section is barns (bn), where 1 bn = $10^{-28}$ $m^2$. If a number of bombarding particles are directed toward a sample of number density n nuclei/$m^3$ and thickness x meters, then the probability of a bombarding particle emerging without having reacted is given by [4]

$$P_{eacape} = e^{-\sigma n x}. \tag{2}$$

In the present case we are concerned with neutrons striking target nuclei in a sample of fissile material. In general, one of three reactions will happen: (i) the incident neutron will induce the struck nucleus to fission (f), (ii) the incident neutron may be absorbed (a) by the struck nucleus, or (iii) the incident neutron may be scattered (s) by the struck nucleus. Each possibility has a corresponding cross section: $\sigma_f$, $\sigma_a$, or $\sigma_s$. For the types of fissile material employed in weapons the absorption cross-sections are usually negligibly small, so we drop that possibility from further consideration. The *total* cross section is then defined as

$$\sigma_{tot} = \sigma_f + \sigma_s. \tag{3}$$

Cross-sections and $\nu$ values are functions of the energy of the bombarding particles; in the case of fission weapons the secondary neutrons are highly energetic (~ 1 MeV or more). For simplicity, I assume that all neutrons are monoenergetic.

Now consider a spherical sample of fissile material of radius R. As nuclei are struck and fission they release secondary neutrons. The question is: How many of these ultimately reach the surface and escape? Neutrons that do escape can do so in one of two ways: (i) they may escape *directly* after being liberated by traveling in a straight line from their birthplace to the edge of the sphere, or (ii) they may scatter one or more times before escaping. To develop the calculation, it is useful to imagine that the maximum possible number of scatterings-before-escape is known in advance: call this $S_{max}$. Exactly how $S_{max}$ is determined is addressed in Section 3.

First consider neutrons that escape directly. The probability of their doing so will be given by Eq. (2) with the total cross-section of Eq. (3) since they avoid both being scattered or consumed in fissioning a nucleus. Assuming that neutrons are emitted in random directions, we can replace the linear distance x in Eq. (2) with the average distance <d> from any point within a sphere to its surface, which can be shown to be <d> = 3R/4. It is convenient to write this as <d> = 3R/4 = $\gamma R$. The number of neutrons that directly escape will then be given by

$$N_{direct\ escape} = \nu N e^{-\xi}, \qquad (4)$$

where

$$\xi = \sigma_{tot} n \gamma R. \qquad (5)$$

Also define the analog of $\xi$ for only the fission cross-section:

$$\xi_f = \sigma_f n \gamma R. \qquad (6)$$

The task now is to add up all neutrons that escape either directly or after suffering one or more scatterings. Let $N_S$ designate the number of neutrons that have survived S scatterings. The number of these that escape following their S'th scattering must be $N_S \exp(-\xi)$ *unless* $S = S_{max}$, in which case the number would be $N_{Smax} \exp(-\xi_f)$. Note the different $\xi$'s in play here. If further scatterings are possible then a neutron must avoid both scattering and fission in order to make its escape, in which case the total cross-section (Eq. 5) applies; if no further scatterings are to be permitted then a neutron has only to avoid being consumed in a fission in order to escape (Eq. 6). The total number of neutrons that escape will be the sum of those that escape directly plus those that escape after S scatterings (for S = 1 to $S_{max}-1$) plus those that escape without being lost to a fission after undergoing $S_{max}$ scatterings:

$$N_{escape} = N_{direct \atop escape} + \sum_{S=1}^{S_{max}-1} N_S e^{-\xi} + N_{S_{max}} e^{-\xi_f}. \tag{7}$$

To develop a general expression for $N_S$ requires tracking of the probabilities of possible fates facing neutrons that do not directly escape.

From Eq. (4) the number of neutrons that do not escape directly must be

$$N_{not} = \nu N(1-e^{-\xi}). \tag{8}$$

As these neutrons do not escape directly they must first interact with a nucleus either by causing a fission or by being scattered. The probabilities of these competing processes will be $P_{fiss} = \sigma_f/\sigma_{tot}$ and $P_{scatt} = \sigma_s/\sigma_{tot}$. Hence the number of neutrons that must suffer at least one scattering must be

$$N_1 = P_{scatt} N_{not} = \nu N \left(\frac{\sigma_s}{\sigma_{tot}}\right)(1-e^{-\xi}). \tag{9}$$

Having suffered a scattering, a neutron now faces *three* fates: (a) it may now escape, (b) it may be consumed in a fission, or (c) it may suffer a subsequent scattering. The probability of escape is $e^{-\xi}$; the *total* cross-section is involved here as both scattering and fission are to be avoided. The probability of not escaping is then $(1-e^{-\xi})$, and so the probability of not escaping and scattering is $(\sigma_s/\sigma_{tot})(1-e^{-\xi})$. Combining this with Eq. (9), the number of neutrons that will scatter at least two or times must be

$$N_2 = \nu N \left(\frac{\sigma_s}{\sigma_{tot}}\right)^2 (1-e^{-\xi})^2. \tag{10}$$

By propagating this reasoning to further scatterings it is easy to see that the number of neutrons that suffer S scatterings will be

$$N_S = (\nu N) g^S \tag{11}$$

where

$$g = \left(\frac{\sigma_s}{\sigma_{tot}}\right)(1-e^{-\xi}). \tag{12}$$

Incorporating Eqs. (4), (11) and (12) into Eq. (7) and recalling the definition of criticality from the Introduction, we can write the critical condition as

$$e^{-\xi}\left(1 + \sum_{S=1}^{S_{max}-1} g^S\right) + g^{S_{max}} e^{-\xi_f} \leq 1 - \frac{1}{\nu}. \tag{13}$$

The summation is a partial sum of a geometric series, which can be expressed in a closed form. This reduces Eq. (13) to

$$e^{-\xi}\left(\frac{1-g^{S_{max}}}{1-g}\right) + g^{S_{max}} e^{-\xi_f} \leq 1 - \frac{1}{\nu}. \tag{14}$$

Eq. (14) is the central result of this paper. It should be noted that we are implicitly assuming that scattering randomly redirects neutrons so that they subsequently face a new mean-distance-to-escape of $\gamma R$ after each scatter.

### 3. Determining $S_{max}$ and the critical radius

Eq. (14) cannot be solved analytically for $\xi$ (or, equivalently, for the critical radius) that will just cause the inequality to be satisfied. There is also the question of what maximum number of scatterings to allow. Indeed, numerical exploration of Eq. (14) reveals that multiple solutions are possible in the sense that it is possible to solve for different values of R upon choosing different values for $S_{max}$. To see how to isolate a unique answer it is helpful to consider more closely the role of scattering.

Two observations on the role of scattering are relevant here. The first is that scattering neutrons are essentially random-walking their way through the fissile material. For a reaction characterized by cross-section $\sigma$, the mean free path between reactions can be shown [4] to be $\lambda = 1/\sigma n$. It is well known that a random walk of S steps each of length $\lambda$ results in an average displacement of $\sqrt{S}\lambda$ from the starting place. In the present case this displacement cannot be allowed to exceed the mean distance to the edge of the sphere or else the neutrons would have scattered out of it: we must have $\sqrt{S_{max}}\lambda_{scatt} \leq \gamma R$. Since $\sqrt{S_{max}}\lambda_{scatt}$ increases while $\gamma R$ decreases with increasing $S_{max}$ (see below) this constraint implies an upper limit on $S_{max}$. To decide from among the various solutions for $S_{max}$ we must choose the one corresponding to the highest value of $S_{max}$ consistent with satisfying $\sqrt{S_{max}}\lambda_{scatt} \leq \gamma R$. This is because choosing a lower value for $S_{max}$ would result in a value of $\sqrt{S_{max}}\lambda_{scatt}$ which in comparison with $\gamma R$ would have permitted more scatterings to have occurred than was assumed, a physically inconsistent situation.

The second observation is that the critical radius will be reduced for larger values of the scattering cross-section. The reason for this is that scattering causes neutrons to travel through more material than if they had made straight-line flights to the edge of the sphere, consequently increasing their chance of causing a fission along the way. This allows us to establish a maximum possible critical radius for any fissile material. If we imagine $\sigma_s = 0$, then one has only to deal with neutrons that escape directly. In this case $g = 0$ and $\xi = \xi_f$, leading to

$$R_{max} = \frac{1}{\sigma_f n \gamma} \ln\left(\frac{\nu}{\nu - 1}\right). \tag{15}$$

If at the same time we are to have $\sqrt{S_{max}}\lambda_{scatt} \leq \gamma R$, then

$$\sqrt{S_{max}} \leq \left(\frac{\sigma_s}{\sigma_f}\right) \ln\left(\frac{\nu}{\nu - 1}\right). \tag{16}$$

If $\sigma_s$ is sufficiently small then not even a single scattering may be permissible, in which case Eq. (15) would apply with $\sigma_f$ replaced by $\sigma_{tot}$.

In practice, I have found that determining a consistent solution proceeds quickly by setting up Eq. (14) in a spreadsheet where the operator can adjust $S_{max}$ and R as desired. The results discussed below were computed with such a spreadsheet, which can be downloaded from http://othello.alma.edu/~reed/CritMass.xls

## 4.  Results

Table 1 shows calculated critical masses for five nuclides usually considered for use in nuclear weapons; in all cases I found $S_{max} = 1$ to be the valid solution. However, this a consequence of my adopting a constant value for the ratio $\sigma_s/\sigma_f$. In general, I was unable to locate figures for scattering cross-sections. However, Wehr et. al. [5] give the fission and total cross-sections of $^{235}_{92}U$ for fast neutrons as 1.3 and 6.5 bn, respectively. Presuming cross-sections for other processes (delayed fissions or non-fission absorptions) to be negligible, we can infer $\sigma_s \sim 5.2$ barn, or $\sigma_s/\sigma_f = 4$, which I adopted for all cases in the Table.

The results obtained here for $^{235}_{92}U$ and $^{239}_{94}Pu$ are in surprisingly good agreement with published values: in the *Los Alamos Primer*, Serber [1] quotes 56 and 11 kg, respectively. One must be careful about making too much of this agreement, however: if $\sigma_f$ for U-235 is increased to 1.3 from Serber's value of 1.22, the critical mass drops to 46 kg (keeping $\sigma_s/\sigma_f = 4$). Nevertheless, it is satisfying that the present simple model gives such apparently credible results.

A surprising result is the small critical mass for Uranium 233. The practical difficulty with this isotope is that it has a relatively short $\alpha$-decay half life ($\sim$ 160,000 yr) and must be synthesized by neutron bombardment of Thorium-232 [5]. The same objection applies for Np-237 and Am-241. However, a Los Alamos news release in October 2002 [6] revealed that researchers there had succeeded in demonstrating Neptunium criticality with a six-kilogram Np sphere in combination with approximately 60 kg of enriched uranium; the Np sphere by itself was of insufficient mass to maintain criticality.

The last line of Table 1 lists spontaneous-fission (SF) rates (per kilogram per microsecond ) for the various nuclides; these were derived from the SF half-lives tabulated in [8]. Such rates are an important consideration in weapons design. The entire fission process happens over a timescale of about one microsecond [1-3]. If within the brief time it takes to assemble subcritical components into a critical circumstance a stray neutron should cause an initial fission, the result is likely to be a fizzle with only a small amount of material being fissioned. The SF rate should be kept to much less than one per microsecond to avoid any significant probability of predetonation. [Indeed, it was "poisoning" of reactor-synthesized Pu-239 by a small amount of Pu-240 (SF rate ~ 0.5 $kg^{-1}$ $\mu s^{-1}$) which rendered the "gun" assembly method unsuitable for the Trinity and Nagasaki plutonium bombs of WW II; only an implosion mechanism could achieve sufficiently fast assembly.] A survey of the *Chart of the Nuclides* and SF half-lives indicates that, due to short alpha or beta half-lives and/or high SF rates, no nuclides beyond those listed in Table 1 are likely to be suitable candidates for weapons materials.

Finally, it is worth remarking that one can arrange to make a weapon with a *lesser* mass of fissile material than has been computed here. The essence of solving Eq. (14) is that for a fixed value of $S_{max}$ the left side of the equation decreases as one tries larger and larger values of R; there is a threshold value of R at which the inequality is just satisfied. However, wherever R appears, it is always multiplied by the number density n; we could just as well increase n until the inequality is satisfied. What is important is the concept of critical density: *any* mass of fissile material can be made into a critical mass if crushed to sufficiently high density. More generally, the criticality condition can be written as $R\rho \geq k$ where k is a constant for a given material. From the numbers in Table 1, $k(^{235}_{92}U) = 168.4$ gr/cm$^2$. The critical mass corresponding to the "normal" density of the material is known as the "bare" critical mass. This radius-density relationship underlies the concept of implosion-initiated weapons. If a sufficiently strong implosion can be achieved then one can get away with using less than a bare critical mass by starting with a hollow-cored sphere of normal density which is implosively crushed to a density great enough to achieve criticality. References [1] and [3] present analyses of to what radius a core comprising a given number of bare critical masses can be allowed to expand before the reaction shuts down. Such analyses, however, are beyond the purpose of the present work.

## 5. Summary

This paper presents a time-independent model for the random-walk scattering of neutrons through a sphere of fissile material; the result is an expression from which critical masses can be estimated. Despite the simplicity of this model it gives results in surprisingly good agreement with published values. A critical sphere of Pu-239 is about the size of a small grapefruit.

## Table 1

## Fissile Nuclides

| Parameter | | $^{233}_{92}U$ | $^{235}_{92}U$ | Nuclide $^{237}_{93}Np$ | $^{239}_{94}Pu$ | $^{241}_{95}Am$ |
|---|---|---|---|---|---|---|
| A | (gr/mol) | 233.04 | 235.04 | 237.05 | 239.05 | 241.06 |
| $\rho$ | (gr/cm$^3$) | 18.95 | 18.95 | 20.25 | 19.84 | 13.67 |
| $\sigma_f$ | (bn) | 1.93 | 1.22 | 1.60 | 1.73 | 1.60 |
| $\sigma_s$ | (bn) | 7.72 | 4.88 | 6.40 | 6.92 | 6.40 |
| $\nu$ | | 2.69 | 2.52 | 2.81 | 2.95 | 2.5 |
| $R_{crit}$ | (cm) | 5.129 | 8.884 | 5.582 | 5.013 | 9.730 |
| $M_{crit}$ | (kg) | 10.7 | 55.7 | 14.8 | 10.5 | 52.7 |
| SF | kg$^{-1}$ µs$^{-1}$ | < 2.1(-7) | 5.6(-9) | < 5.6(-8) | 6.9(-6) | 4.6(-4) |

Notes and sources for Table 1:

Atomic weights (A) and bulk densities ($\rho$) were adopted from the *Chart of the Nuclides* [9]. All nuclides are assumed to have $\sigma_s/\sigma_f = 4$. In the spontaneous fission (SF) rates, the number in brackets gives the power of ten, e.g., 2.1(-7) = 2.1 x 10$^{-7}$.

U-233: $\sigma_f$ from Hyde [10] p. 72, interpolated for 1 MeV neutrons. $\nu$ from Hyde [10] p. 212 for effective neutron energy 1.3 MeV.

U-235: $\sigma_f$ and $\nu$ from Serber [1] p. 20.

Np-237: $\sigma_f$ from Oblozinsky [11], interpolated for 1 MeV neutrons. $\nu$ from Hyde [10] p. 215 for average neutron energy 1.4 MeV.

Pu-239: $\sigma_f$ and $\nu$ from Serber [1] p. 20.

Am-241: $\sigma_f$ from Kawano [12], interpolated for 1 MeV neutrons. $\nu$ assumed.